# Dynamical decoherence of the light induced interlayer coupling in YBa$_2$Cu$_3$O$_{6+\delta}$

C. R. Hunt[1,4], D. Nicoletti[1], S. Kaiser[1,3], D. Pröpper[3], T. Loew[3], J. Porras[3], B. Keimer[3], A. Cavalleri[1,2]

[1] Max Planck Institute for the Structure and Dynamics of Matter, Hamburg, Germany
[2] Department of Physics, Oxford University, Clarendon Laboratory, Oxford, United Kingdom
[3] Max Planck Institute for Solid State Research, Stuttgart, Germany
[4] Department of Physics, University of California Berkeley, Berkeley, California, USA

**Optical excitation of apical oxygen vibrations in YBa$_2$Cu$_3$O$_{6+\delta}$ has been shown to enhance its *c*-axis superconducting-phase rigidity, as evidenced by a transient blue shift of the equilibrium inter-bilayer Josephson plasma resonance. Surprisingly, a transient *c*-axis plasma mode could also be induced above $T_c$ by the same apical oxygen excitation, suggesting light activated superfluid tunneling throughout the pseudogap phase of YBa$_2$Cu$_3$O$_{6+\delta}$. However, despite the similarities between the above $T_c$ transient plasma mode and the equilibrium Josephson plasmon, alternative explanations involving high mobility quasiparticle transport should be considered. Here, we report an extensive study of the relaxation of the light-induced plasmon into the equilibrium incoherent phase. These new experiments allow for a critical assessment of the nature of this mode. We determine that the transient plasma relaxes through a collapse of its coherence length rather than its carrier (or superfluid) density. These observations are not easily reconciled with quasiparticle interlayer transport, and rather support transient superfluid tunneling as the origin of the light-induced interlayer coupling in YBa$_2$Cu$_3$O$_{6+\delta}$.**

Underdoped cuprates retain some of the properties associated with superconductivity above the transition temperature $T_c$.[1,2,3,4,5,6,7] For instance, unlike in conventional superconductors in which the transition to the normal state is governed by a reduction in condensate density, superconductivity in cuprates is also weakened by fluctuations of the order parameter phase. These fluctuations set in below $T_c$[8,9,10] and persist above the transition temperature.[11,12,13,14] Competing orders—such as the charge and spin stripes found in lanthanides[15,16,17,18,19] and other charge density wave states observed across a wide range of cuprates[20,21,22,23] —may disrupt superconductivity through order parameter modulation.[24,25,26,27]

Targeted ultrafast light excitation has been recently shown to transiently restore superconducting-like properties above equilibrium $T_c$. This is most directly evidenced by the appearance of plasma modes reminiscent of intrinsic Josephson coupling between the $CuO_2$ layers.[28,29,30,31,32,33,34,35,36] In a first set of experiments, light-induced interlayer coupling was observed in single layer lanthanide $La_{1.8-x}Eu_{0.2}Sr_xCuO_4$ after excitation resonant to a Cu-O stretching mode.[37] Because this effect correlates with the static charge order[38] and because the same excitation was also shown to disrupt stripe order in the related compound $La_{0.1875}Ba_{0.125}CuO_4$,[39] transient appearance of superconductivity can probably be explained in 1/8 doped single layer cuprates as a result of competing order melting.

More puzzling has been the response of underdoped $YBa_2Cu_3O_{6+\delta}$,[40,41] for which excitation of the apical oxygen atoms, sitting above and below the $CuO_2$ planes (see Fig. 1A), also promotes a superconducting-like plasma mode. Whereas near 1/8 hole doping ($YBa_2Cu_3O_{6.65}$) the effect was correlated with suppression of charge density wave order,[42] measurements at lower dopings ($YBa_2Cu_3O_{6.5}$ and $YBa_2Cu_3O_{6.45}$) generated light-induced coherence at even higher temperatures. Because equilibrium charge order is weak for these doping levels, the results in $YBa_2Cu_3O_{6+\delta}$ are difficult to reconcile with a competing order melting, and also raise questions on the nature of the plasma mode itself.

Provided that the plasma edge can be uniquely associated with Josephson tunneling also for all underdoped $YBa_2Cu_3O_{6+\delta}$ samples, these experiments indicate that light induced superconductivity may result from a more general mechanism.[43,44] For example, a connection may be drawn with the results of broadband infrared spectroscopy on several bilayer cuprates[5,7], which suggest that local, coherent pair tunneling may survive far above $T_c$ (up to ~200 K in $YBa_2Cu_3O_{6+\delta}$) between the more closely spaced $CuO_2$ planes making up the bilayers within each unit cell. Photo-stimulation may then be enhancing these pre-existing fluctuations through some dynamical mechanism, stabilizing interlayer coherence.

Here we present a comprehensive study of the light-induced interlayer coupling and especially its relaxation in $YBa_2Cu_3O_x$ (YBCO $x$). Five different compounds were studied, consisting of underdoped $x = 6.3, 6.45, 6.5, 6.6$, and optimally doped $x = 7$, with $T_c \simeq 0$ K, 35 K, 50 K, 62 K, and 90 K respectively. The single crystals used in the experiment[40] had typical dimensions of 2 x 2 x 1 mm³. Their $T_c$ values were determined by dc magnetization measurements (see Fig. 1B). YBCO 6.3 was found to be non-superconducting down to the lowest measured temperature.

We photoexcited YBCO with ~300 fs mid-infrared pulses, polarized perpendicular to the $CuO_2$ planes (along the $c$-axis) and with field strengths up to ~3 MV/cm. These pump pulses were tuned to 15 $\mu$m wavelength, to resonantly drive a phonon mode of B$_{1u}$ symmetry, involving motion of the apical oxygen atoms (see Fig. 1A).[40,41,43] The transient optical response was interrogated using delayed single-cycle terahertz pulses with spectral bandwidth covering the 0.5 – 2.5 THz range. The experiment was performed in reflection geometry, with the pump beam striking the sample at normal incidence and the THz probe at 30°. The reflected probe field in the absence of excitation, $E(t)$, and the pump-induced changes to the field at each time delay $\tau$, $\Delta E(t, \tau)$, were measured via electro-optic sampling and then independently Fourier transformed to retrieve the complex response functions of the photo-stimulated material (see Supplementary Section S1).

The same time-domain setup was also used, in the absence of a pump beam, to determine the THz response of YBCO in its equilibrium superconducting state. The normalized difference $|\Delta \tilde{E}(\omega)/\tilde{E}(\omega)|$ between the $c$-axis reflection coefficient at 5 K and that at $T \gtrsim T_c$ was measured for all underdoped samples. This response is dominated by Cooper pair tunneling, as the reflectivity above $T_c$ is flat and featureless in this frequency range at all dopings (see Supplementary Section S2). As shown in the top row of Fig. 1C, a sharp edge appears for $x = 6.45, 6.5, 6.6$ at the Josephson plasma frequency $\omega_p$. The edge position blue shifts with increasing doping $x$, and no mode is found for non-superconducting YBCO 6.3.

The pump-induced changes in reflectivity, $\Delta R(\omega, \tau = 0.8 \text{ ps})/R(\omega)$ are displayed in the same frequency range in the bottom row of Fig. 1C, for all dopings at several temperatures above $T_c$. These were calculated by taking into account the pump-probe penetration depth mismatch,[37,38,40,41] as discussed in Supplementary Section S1. In analogy with the equilibrium superconducting state, the photo-excited state is characterized by the appearance of a reflectivity edge, whose frequency blue shifts with increasing doping, tracking the position of the equilibrium Josephson plasma resonance at $T \ll T_c$. In YBCO 6.3 only a weak upturn in $\Delta R/R$ was found at low frequencies, suggesting that a mode may be appearing outside of the probed spectral range (see later discussion).

The transient response of the light-induced state could be fitted by assuming an effective medium[40,41] consisting of a volume $1 - f$ that retains the optical properties of the material at equilibrium (with complex dielectric function $\tilde{\varepsilon}_{eq}$), and a volume fraction $f$ with the response of a single plasma mode, described as $\tilde{\varepsilon}_p = \tilde{\varepsilon}_c - 4\pi \tilde{\sigma}_p/i\omega$, where,

$$\tilde{\sigma}_p = \frac{1}{4\pi}\left(\frac{\omega_p^2}{\Gamma - i\omega}\right). \quad [1]$$

The $\Gamma$ term encompasses all scattering and decoherence processes affecting the transient plasma mode. The effective response $\tilde{\varepsilon}_E$ is given by the Bruggeman equation,

$$f\left(\frac{\tilde{\varepsilon}_p - \tilde{\varepsilon}_E}{\tilde{\varepsilon}_p + 2\tilde{\varepsilon}_E}\right) + (1-f)\left(\frac{\tilde{\varepsilon}_{eq} - \tilde{\varepsilon}_E}{\tilde{\varepsilon}_{eq} + 2\tilde{\varepsilon}_E}\right) = 0. \quad [2]$$

Effective medium fits to the THz reflectivity are shown as dashed lines in Fig. 1C.

Associated with the appearance of a plasma mode in reflectivity is an increase in imaginary conductivity, $\sigma_2(\omega)$, which for dissipationless transport should diverge like $1/\omega$ for $\omega \to 0$. In Fig. 2 we report the real (middle row) and imaginary part (bottom row) of the optical conductivity measured at equilibrium and at $\tau = 0.8$ ps after excitation. The pump induced changes to the imaginary conductivity, $\Delta\sigma_2(\omega) = \sigma_2(\omega) - \sigma_{2,eq}(\omega)$, are also shown in the top row of the same figure. At all dopings $x \leq 6.5$, $\sigma_2(\omega)$ exhibits an enhancement toward low frequencies, while $\sigma_1(\omega)$ shows little change and remains gapped, as expected for the purely inductive response of a perfect conductor. In YBCO 6.6, where the hole concentration is higher, a small, frequency independent increase in $\sigma_1(\omega)$ is observed, which we attribute to incoherent quasiparticle excitation.[45] This contribution also reshapes $\sigma_2(\omega)$. Dashed lines represent here the same effective medium fits applied in Fig. 1C.

While signatures of light-induced coherence were already reported for YBCO 6.45-6.6,[40] here we show similar evidence also in YBCO 6.3, which is never a superconductor at equilibrium. As in the other compounds, this response is identified through a positive, $1/\omega$-type contribution to $\sigma_2(\omega)$, following Eq. 1 in the limit $\Gamma \to 0$. The associated reflectivity edge is not clearly visible, as it remains below our THz probe spectral window (see Supplementary Section S3 for extended data sets).

Optimally doped YBCO 7 was the only sample for which no light-induced increase in $\sigma_2(\omega)$ could be found at any temperature (right column of Fig. 2). The pump-induced *decrease* in both $\sigma_1(\omega)$ and $\sigma_2(\omega)$ observed in this compound is consistent with quasiparticle heating and supports a scenario in which coherent interlayer transport can only be induced in the pseudogap phase of YBCO. We also note that the 15 $\mu$m excitation wavelength only targets the apical oxygen mode on sites with a chain oxygen vacancy (the YBCO 6 structure) and therefore the optical excitation is detuned from the phonon resonance in optimally doped YBCO 7.[43,46] In Supplementary Section S4 we also discuss the absence of any light-induced coherence in underdoped YBCO after stimulation with 15 μm pulses polarized parallel to the $CuO_2$ planes (along the *a* direction), resonant to an in-plane phonon mode.

In a superconductor at equilibrium, the superfluid density is proportional to the frequency-independent quantity $\omega\sigma_2(\omega)|_{\omega\to 0}$. Along the *c*-axis this response, approximated as $\omega\Delta\sigma_2(\omega) = \omega[\sigma_2(\omega, T < T_c) - \sigma_2(\omega, T > T_c)]$, measures the component of the superfluid that contributes to interlayer tunneling (proportional to $\omega_p^2$), and can be used to quantify the *c*-axis Josephson coupling strength (see Fig. 3A.1). Here we take $\omega\Delta\sigma_2(\omega, \tau)$ as a measure of light-induced interbilayer coherence. This quantity, measured at the peak of the response, is plotted in Fig. 3A.2 for YBCO 6.45 at different temperatures. Its mean value, $\langle\omega\Delta\sigma_2(\omega)\rangle$, was determined across all dopings by averaging $\omega\Delta\sigma_2(\omega)$ in the range where it remained frequency-independent ($\omega \lesssim 1.8$ THz). The temperature dependence of $\langle\omega\Delta\sigma_2(\omega)\rangle$ is displayed in Fig. 3B-E for all compounds, along with the photo-susceptible volume fraction, $f$, extracted from the effective medium fits (see Eq. 2). Due to the inhomogeneous nature of the excitation, the measured $\langle\omega\Delta\sigma_2(\omega)\rangle$ is rescaled by $f$ at all measured dopings.

Two distinct temperature regimes are observed for $\langle\omega\Delta\sigma_2(\omega)\rangle$. Just above $T_c$, this quantity increases with temperature up to a crossover point $T''$. Above $T''$ the response follows the mean field behavior typical of a superfluid, $\langle\omega\Delta\sigma_2(\omega)\rangle \propto \sqrt{1 - T/T'}$, dropping to zero at a temperature $T'$. The $T''$ crossover temperature is estimated by the intersection of a linear fit to the low-temperature $\langle\omega\Delta\sigma_2(\omega)\rangle$ (grey line) and the mean field fit to the high-temperature $\langle\omega\Delta\sigma_2(\omega)\rangle$ (black dashed line).

These two temperature scales are plotted on the phase diagram in Fig. 3F. The similarity between $T'$ and the $T^*$ line associated with the pseudogap phase is apparent. The $T''$ crossover appears to track the $T_{ELC}$ scale, identified in Refs. 47,48 as the electronic liquid crystal (ELC) temperature, where there is an onset of nematic behavior due to collective excitations between spins. The reduction of $\langle\omega\Delta\sigma_2(\omega)\rangle$ below this temperature suggests that the ELC ground state may be competing with the light-induced phase.

We now focus on the relaxation dynamics of the transient *c*-axis coupling, which is characterized by a drop in the imaginary conductivity below a frequency $\omega^*$ along with a decrease in the overall magnitude of $\omega\Delta\sigma_2(\omega)$. The time evolution of $\omega\Delta\sigma_2(\omega)$ after excitation is shown in Fig. 4A for YBCO 6.45 at 100 K, while extended data sets are reported in Supplementary Section S5. The deviation from a frequency-independent behavior for $\omega\Delta\sigma_2(\omega)$ indicates an effective scattering rate ($\omega^* \cong \Gamma$ in Eq. 1) for the interlayer coupling which increases in time, corresponding to a decrease of the coherence length along the *c* axis. This is reminiscent of the $\omega\sigma_2(\omega)$ measured in $La_{2-x}Sr_xCuO_4$[9] and $Bi_2Sr_2CaCu_2O_{8+\delta}$[11] near $T_c$, which was ascribed to fluctuations in the superconducting correlation length and time scale.

The coherence length $d$ of the transient state can be expressed as $d = 2\omega_p L/\omega^*$, where $L$ is the interbilayer spacing. This is plotted as a function of time delay in Fig. 4B. The drop in coherence follows a double exponential fit, with a fast ($\tau_1 \lesssim 1$ ps) and a slow ($\tau_2 \simeq 4$ ps) relaxation component. The overall magnitude of $\langle\omega\Delta\sigma_2(\omega)\rangle$, measured for $\omega > \omega^*$, also follows a similar decay, as shown in Fig. 4C. The temperature dependence of the $\langle\omega\Delta\sigma_2(\omega)\rangle$ relaxation timescales are displayed in Fig. 4D. Remarkably, $\tau_2$ increases with increasing temperature until a turning point near $T'' \approx 150$ K, and then drops off, preserving a small but finite value even at the highest temperature measured.

During the relaxation, a second distinct phenomenon emerges. The light-induced plasma resonance splits into two closely-spaced modes. The splitting is most clearly visualized in the energy loss function, $LF(\omega) = -\text{Im}[1/\tilde{\varepsilon}(\omega)]$, displayed in Fig. 5 for YBCO 6.45 at 200 K. At τ = 0.8 ps, the light-induced plasma mode produces a single peak in the loss function, centered near $\omega_p$, while at τ = 1.8 ps, part of the spectral weight has blue-shifted, resulting in two peaks separated by ~1 THz. Correspondingly, $\sigma_1(\omega)$ develops a peak at an intermediate frequency.

Spatial variations of the plasma resonance frequency, for example due to inhomogeneities, would simply result in a broadening of the plasmon.[32] The appearance of two distinct modes is instead indicative of the development of two inequivalent interbilayer couplings. A theoretical description of multiple, inequivalent junctions was developed in Refs. 32,33 to account for the bilayer structure of $YBa_2Cu_3O_{6+\delta}$. We employ this same model for the split resonance observed during the relaxation dynamics, resulting in a complex dielectric function $\tilde{\varepsilon}_{2p}$, described by the expression:

$$\frac{\varepsilon_\infty}{\tilde{\varepsilon}_{2p}} = \frac{\omega^2 \tilde{z}_1}{\omega^2 - \omega_{p1}^2 + i\omega\Gamma_1} + \frac{\omega^2 \tilde{z}_2}{\omega^2 - \omega_{p2}^2 + i\omega\Gamma_2}, \quad [3]$$

Here $\omega_{p1}$ and $\omega_{p2}$ are the two plasma frequencies and $\Gamma_{1,2}$ capture incoherent contributions to the response. The relative weights of each mode, $\tilde{z}_{1,2}$, are related to the geometry and electronic compressibility of each junction.[33] Effective medium fits, performed substituting $\tilde{\varepsilon}_p$ in Eq. 2 with $\tilde{\varepsilon}_{2p}$, are shown as dashed lines in Fig. 5. The two responses used in the effective medium, volume $f$ with response $\tilde{\varepsilon}_{2p}$ and $1 - f$ with the equilibrium response $\tilde{\varepsilon}_{eq}$, are shown in Fig. 5C. The peak in $\sigma_1(\omega)$ corresponds to a "transverse" plasma mode,[33] centered at frequency $\omega_T = \sqrt{\tilde{z}_1 \omega_{p2}^2 + \tilde{z}_2 \omega_{p1}^2}$. This splitting of the transient plasmon is a feature of the relaxation for all dopings, at all temperatures where the mode could be generated (see Supplementary Section S6).

A split Josephson plasmon could be caused by the generation of phase slips or vortices during relaxation, in a manner similar to the thermal vortex regime that forms an extended dome above $T_c$ in the equilibrium phase diagram of cuprates.[8,11,49,50] Vortices produced by an applied magnetic field along the $CuO_2$ planes have been shown to induce a splitting of the equilibrium interbilayer Josephson plasmon in a similar fashion.[34,35,51] The spatial distribution of the phase variation determines the degree and shape of the splitting.[36]

The most natural implication of all results reported above involves transient light-induced superconducting fluctuations at unprecedented high temperatures.[40,41] This possibility has been raised in previous papers and the data reported here provide further support to this view, especially in relation to an alternative interpretation that posits the formation of a plasma of high mobility quasiparticles. Note that such a quasiparticle plasma would have to be at the same frequency of the Josephson plasma edge below $T_c$, with the observed low scattering rates ($\Gamma \lesssim 0.5$ THz) that are difficult to explain in terms of interlayer hopping of incoherent carriers.

Most importantly, the time evolution of the transient plasmon points to decoherence as an important driver of the relaxation to the ground state. This occurs through a collapse of the coherence length $d$ (see Fig. 4) or, equivalently, via an increase in the effective scattering rate $\Gamma$ (see Eq. 1). In contrast, the response of Drude quasiparticles following photo-excitation in doped semiconductors typically results in a relaxation through a depletion of the carrier density at constant scattering time,[52,53] or with a *decrease* in the scattering rate.[54,55] Finally, the observed splitting of the plasma mode, which fits nicely in a picture of a superfluid developing phase slips, cannot be explained by a plasma of mutually incoherent quasiparticles.

In summary, we have reported on an extensive study of the light-induced coherent interlayer coupling in $YBa_2Cu_3O_{6+\delta}$ throughout its pseudogap phase. Both the amplitude and lifetime of the response indicate the existence of a temperature scale $T''$ below which the transient plasmon is suppressed. This temperature coincides with the onset of collective spin excitations, suggesting that spin ordering may depress light-induced coherence. Furthermore, the relaxation of the light-induced state appears to follow a dynamics that is well aligned with the loss of phase coherence in the tunneling of a superfluid through the layers. This observation further supports an analogy with transient superconducting fluctuations. One challenge for future work is to extend the lifetime of the transient state, supporting the interlayer coupling against decoherence effects.

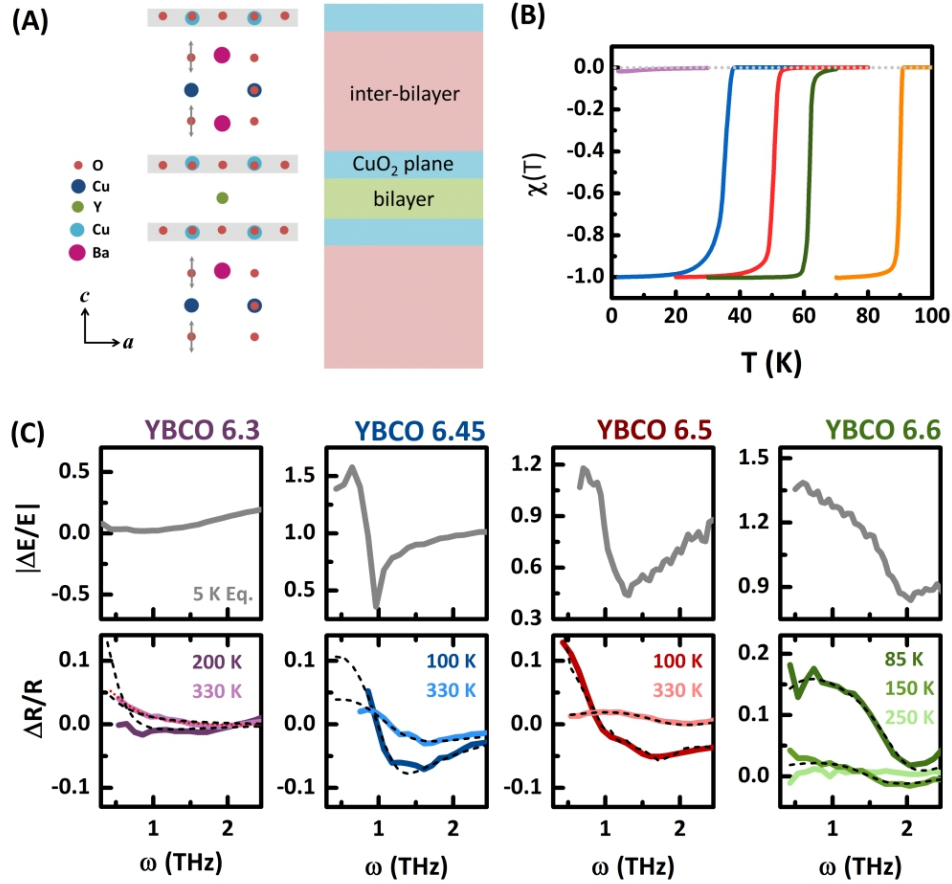

**Figure 1. Equilibrium superconducting response of YBCO and transient THz reflectivity. (A)** Crystal structure of YBCO. Grey arrows indicate the apical oxygen excitation. The unit cell can be divided into two regions, the bilayer units and the interbilayer gaps, separated by $CuO_2$ planes. **(B)** Equilibrium magnetic susceptibility of the five samples of $YBa_2Cu_3O_x$, with $x = 6.3$ (purple), 6.45 (blue), 6.5 (red), 6.6 (green), and 7 (orange). These correspond to hole dopings of 0.05, 0.07, 0.09, 0.12, and 0.16, respectively. The $x = 6.3$ sample is non-superconducting, while the other samples have transition temperatures of $T_c$ = 35 K, 51 K, 62 K, and 90 K, respectively. **(C, top row)** Change in the equilibrium THz reflection coefficient upon cooling below $T_c$. Sharp edges are found in the superconducting samples at the Josephson plasma resonance frequency. **(C, bottom row)** Light-induced reflectivity changes measured for different $T > T_c$, at 0.8 ps after excitation. Fits with an effective medium model (see main text) are displayed as dashed lines.

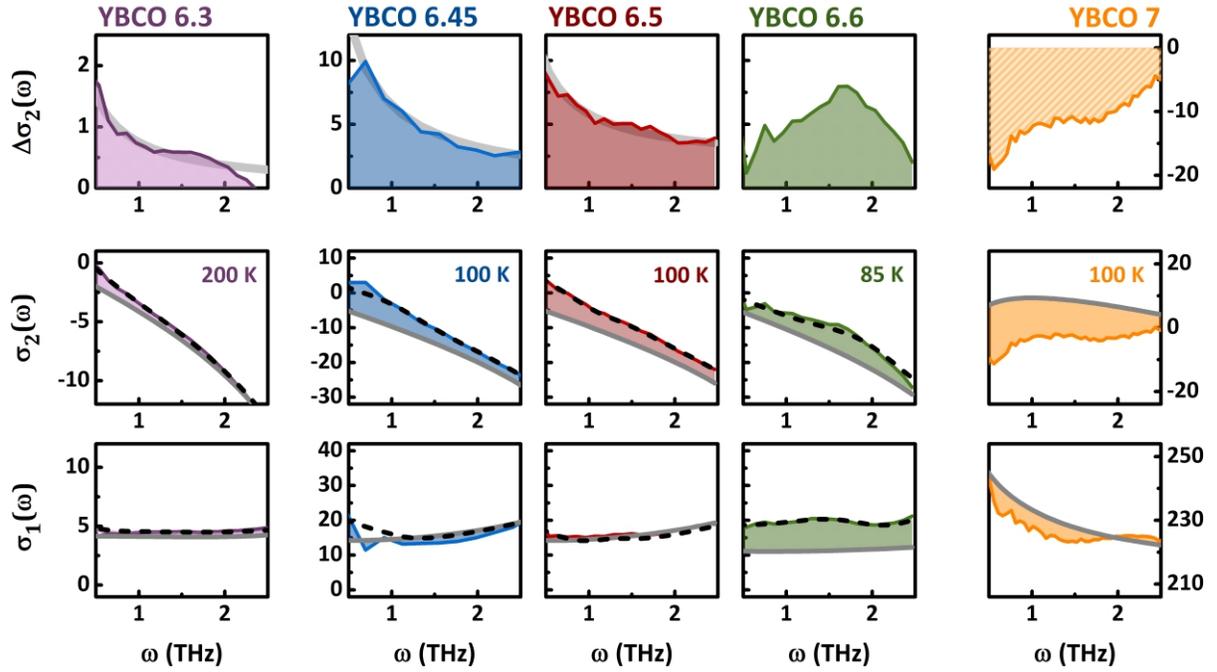

**Figure 2. Transient complex optical conductivity. (Top row)** Pump-induced changes to the imaginary conductivity $\Delta\sigma_2(\omega) = \sigma_2(\omega, \tau = 0.8 \text{ ps}) - \sigma_{2,eq}(\omega)$. Grey lines indicate $1/\omega$ fits. A quasiparticle contribution reshapes $\Delta\sigma_2(\omega)$ in YBCO 6.6. Corresponding imaginary **(middle row)** and real part **(bottom row)** of the optical conductivity at equilibrium (grey) and at $\tau = 0.8$ ps (colored by doping), expressed in units of $\Omega^{-1}\text{cm}^{-1}$. Shaded regions highlight the pump-induced changes. Fits with an effective medium model are shown as dashed lines.

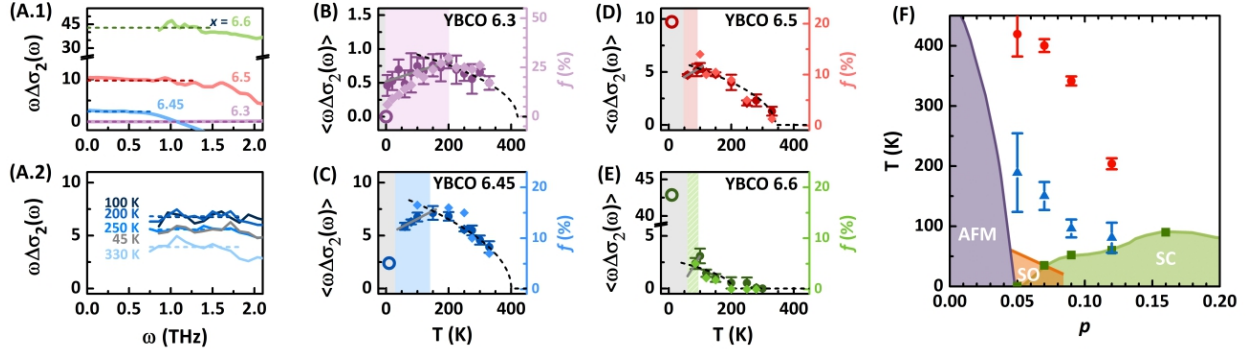

**Figure 3. Phase diagram of the transient state above $T_c$. (A.1)** Equilibrium changes in $\omega\sigma_2(\omega)$ across the superconducting transition, expressed in units THz/($\Omega \cdot$ cm). **(A.2)** Light-induced $\omega\Delta\sigma_2(\omega, \tau = 0.8\ \text{ps})$ measured in YBCO 4.5 at five different temperatures. Dashed lines indicate the mean values, $\langle\omega\Delta\sigma_2(\omega)\rangle$, extracted to quantify the strength of interbilayer coupling (see main text). **(B-D)** Frequency-averaged $\langle\omega\Delta\sigma_2(\omega)\rangle$, plotted as a function of temperature (filled circles). Equilibrium $\langle\omega\Delta\sigma_2(\omega)\rangle$ values measured at 5 K, are also shown (empty circles). Diamonds refer to the photo-susceptible volume fractions, $f$, extracted from effective medium fits. The region $T < T''$ is indicated by a colored background for each doping. Linear fits to $\langle\omega\Delta\sigma_2(\omega)\rangle$ are displayed in this region as grey lines. At $T > T''$, $\langle\omega\Delta\sigma_2(\omega)\rangle$ is well reproduced by a mean field behavior of the type $\propto \sqrt{1 - T/T'}$ (black dashed lines). **(E)** $T'$ (red circles) and $T''$ (blue triangles) values, extracted from the fits, are plotted on the YBCO phase diagram. The equilibrium superconducting transition $T_c$, determined by dc magnetization, is denoted by green squares.

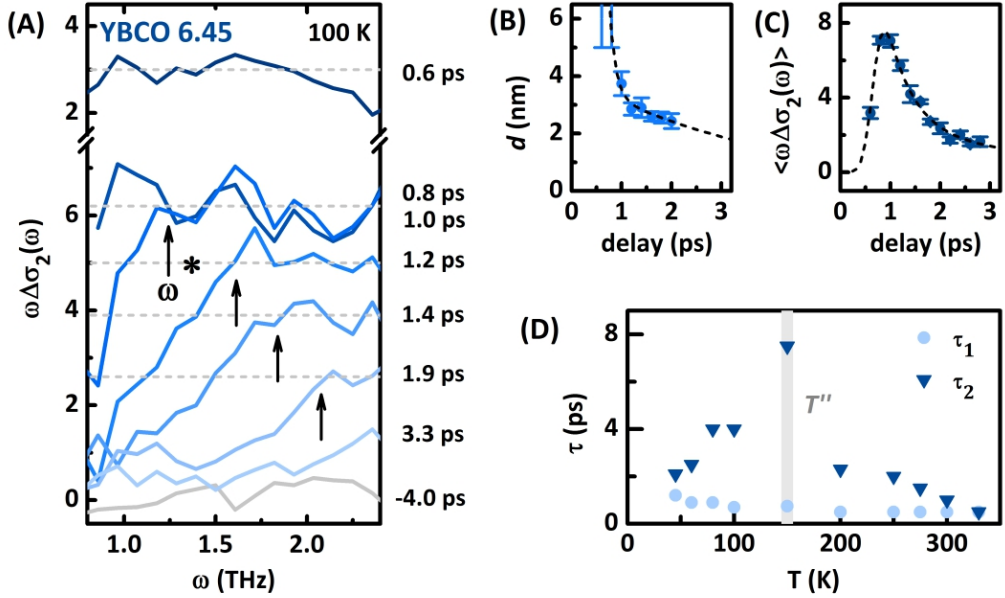

**Figure 4. Decoherence driven relaxation of the transient state. (A)** Transient $\omega\Delta\sigma_2(\omega)$ measured in YBCO 6.45 at $T$ = 100 K, for different pump-probe delays. A low-frequency downturn appears below a characteristic frequency $\omega^*$. **(C)** Relaxation of the coherence length $d = 2\omega_p L/\omega^*$ (see main text) of the transient high mobility state. A double exponential fit is displayed as dashed line. **(D)** Corresponding relaxation of $\langle\omega\Delta\sigma_2(\omega)\rangle$, showing similar time constants. **(E)** Relaxation time constants of $\langle\omega\Delta\sigma_2(\omega)\rangle$, plotted as a function of temperature. The lifetime peaks near $T''$.

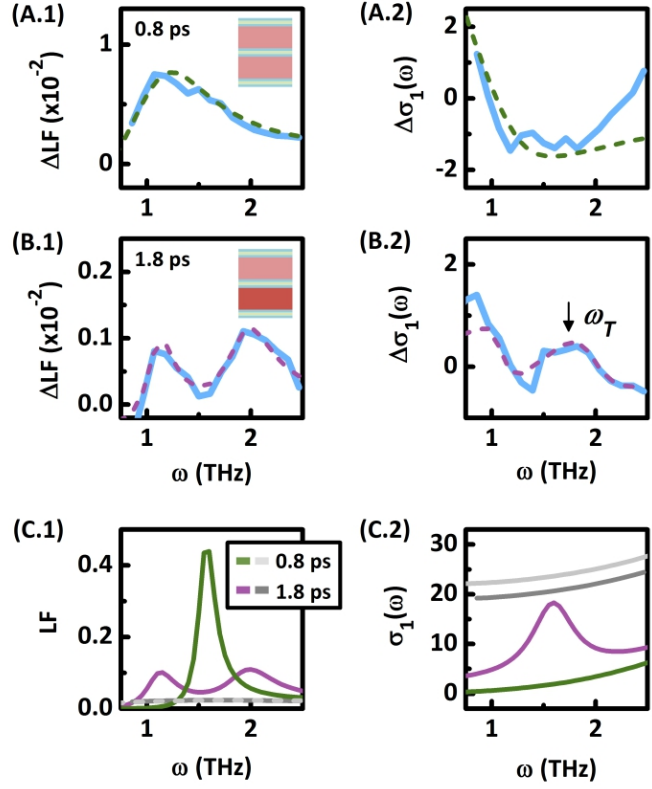

**Figure 5. Splitting of the plasma resonance during relaxation. (A-B)** Pump-induced changes in the loss function and in the real part of the optical conductivity, measured in YBCO 6.45 at $T$ = 200 K at two selected time delays. The splitting of the plasma mode **(B.1)** and the appearance of a peak in $\sigma_1(\omega)$ **(B.2)**, observed at $\tau = 1.8$ ps, indicate that two inequivalent interbilayer junctions have formed. Effective medium fits to the data (see main text) are shown as dashed lines. **(C)** Loss function and real conductivity of the two effective medium components, as extracted from the fits.

**References**

[1] T. Timusk and B. Statt, "The pseudogap in high-temperature superconductors: an experimental survey," *Rep. Prog. Phys.* **62**, 61 (1999).

[2] Z. A. Xu, N. P. Ong, Y. Wang, T. Kakeshita, and S. Uchida, "Vortex-like excitations and the onset of superconducting phase fluctuation in underdoped $La_{2-x}Sr_xCuO_4$," *Nature* **406**, 486 (2000).

[3] D. N. Basov and T. Timusk, "Electrodynamics of high-$T_c$ superconductors," *Rev. Mod. Phys.* **77**, 721 (2005).

[4] M. Hücker, M. von Zimmermann, G. D. Gu, Z. J. Xu, *et al.*, "Stripe order in superconducting $La_{2-x}Ba_xCuO_4$ ($0.095 \leq x \leq 0.155$)," *Phys. Rev. B* **83**, 104506 (2011).

[5] A. Dubroka, M. Rössle, K. W. Kim, V. K. Malik, D. Munzar, D. N. Basov, A. A. Schafgans, S. J. Moon, C. T. Lin, D. Haug, V. Hinkov, B. Keimer, Th. Wolf, J. G. Storey, J. L. Tallon, and C. Bernhard. "Evidence of a Precursor Superconducting Phase at Temperatures as High as 180 K in $RBa_2Cu_3O_{7-x}$ (R=Y,Gd,Eu) Superconducting Crystals from Infrared Spectroscopy," *Phys. Rev. Lett.* **106**, 047006 (2011).

[6] Q. Li, M. Hücker, G. D. Gu, A. M. Tsvelik, and J. M. Tranquada, "Two-dimensional superconducting fluctuations in stripe-ordered $La_{1.875}Ba_{0.125}CuO_4$," *Phys. Rev. Lett.* **99**, 067001 (2007).

[7] E. Uykur, K. Tanaka, T. Masui, S. Miyasaka, and S. Tajima, "Persistence of the Superconducting Condensate Far above the Critical Temperature of $YBa_2(Cu,Zn)_3O_y$ Revealed by *c*-Axis Optical Conductivity Measurements for Several Zn Concentrations and Carrier Doping Levels," *Phys. Rev. Lett.* **112**, 127003 (2014).

[8] V. J. Emery and S. A. Kivelson, "Importance of phase fluctuations in superconductors with small superfluid density, " *Nature* **374**, 434 (1995).

[9] L. S. Bilbro, R. Valdés Aguilar, G. Logvenov, O. Pelleg, I. Bozovic, and N. P. Armitage, "Temporal correlations of superconductivity above the transition temperature in $La_{2-x}Sr_xCuO_4$ probed by terahertz spectroscopy," *Nat. Phys.* **7**, 298–302 (2011).

[10] D. Nakamura, Y. Imai, A. Maeda, and I. Tsukada, "Superconducting fluctuation investigated by THz conductivity of $La_{2-x}Sr_xCuO_4$ thin films," *J. Phys. Soc. Jpn.* **81,** 044709 (2012).

[11] J. Corson, R. Mallozzi, J. Orenstein, J. N. Eckstein, and I. Bozovic, "Vanishing of phase coherence in underdoped $Bi_2Sr_2CaCu_2O_{8+\delta}$," *Nature* **398**, 221 (1999).

[12] C. Stock, W. J. L. Buyers, R. Liang, D. Peets, Z. Tun, D. Bonn, W. N. Hardy, and R. J. Birgeneau. "Dynamic stripes and resonance in the superconducting and normal phases of $YBa_2Cu_3O_{6.5}$ ortho-II superconductor," *Phys. Rev. B* **69**, 014502 (2004).

[13] Y. Wang, L. Li, and N. P. Ong, "Nernst effect in high-$T_c$ superconductors," *Phys. Rev. B* **73**, 024510 (2006).

[14] L. Li, Y. Wang, S. Komiya, S. Ono, Y. Ando, G. D. Gu, and N. P. Ong, "Diamagnetism and Cooper pairing above $T_c$ in cuprates," *Phys. Rev. B* **81**, 054510 (2010).

[15] J. M. Tranquada, B. J. Sternlieb, J. D. Axe, Y. Nakamura, and S. Uchida, "Evidence for stripe correlations of spins and holes in copper oxide superconductors," *Nature* **375**, 561 (1995).

[16] V. J. Emery, S. A. Kivelson, and J. M. Tranquada, "Stripe phases in high-temperature superconductors," *Proc. Nat. Acad. Sci.* **96**, 8814 (1999).

[17] M. Fujita, H. Goka, K. Yamada, and M. Matsuda, "Competition between charge- and spin-density-wave order and superconductivity in $La_{1.875}Ba_{0.125-x}Sr_xCuO_4$," *Phys. Rev. Lett.* **88**, 167008 (2002).

[18] H. H. Klauss, W. Wagener, M. Hillberg, W. Kopmann, H. Walf, F. J. Litterst, M. Hücker, and B. Büchner. "From antiferromagnetic order to static magnetic stripes: The phase diagram of $(La,Eu)_{2-x}Sr_xCuO_4$," *Phys. Rev. Lett.* **85**, 4590 (2000).

[19] M. Hücker, G. D. Gu, J. M. Tranquada, M. von Zimmermann, H.-H. Klauss, N.J. Curro, M. Braden, and B. Büchner. "Coupling of stripes to lattice distortions in cuprates and nickelates," *Physica C* **460**, 170 (2007).

[20] G. Ghiringelli, M. L. Tacon, M. Minola, S. Blanco-Canosa, C. Mazzoli, N. B. Brookes, G. M. De Luca, A. Frano, D. G. Hawthorn, F. He, T. Loew, M. Moretti Sala, D. C. Peets, M. Salluzzo, E. Schierle, R. Sutarto, G. A. Sawatzky, E. Weschke, B. Keimer, and L. Braicovich. "Long-range incommensurate charge fluctuations in $(Y,Nd)Ba_2Cu_3O_{6+x}$," *Science* **337**, 821 (2012).

[21] E. H. da Silva Neto, P. Aynajian, A. Frano, R. Comin, E. Schierle, E. Weschke, A. Gyenis, J. Wen, J. Schneeloch, Z. Xu, S. Ono, G. Gu, M. Le Tacon, and A. Yazdani. "Ubiquitous interplay between charge ordering and high-temperature superconductivity in cuprates," *Science* **343**, 393 (2014).

[22] R. Comin, A. Frano, M. M. Yee, Y. Yoshida, H. Eisaki, E. Schierle, E. Weschke, R. Sutarto, F. He, A. Soumyanarayanan, Y. He, M. Le Tacon, I. S. Elfimov, J. E. Hoffman, G. A. Sawatzky, B. Keimer, and A. Damascelli. "Charge Order Driven by Fermi-Arc Instability in $Bi_2Sr_{2-x}LaxCu_2O_{6+\delta}$," *Science* **343**, 390 (2014).

[23] W. Tabis, Y. Li, M. L. Tacon, L. Braicovich, A. Kreyssig, M. Minola, G. Dellea, E. Weschke, M. J. Veit, M. Ramazanoglu, A. I. Goldman, T. Schmitt, G. Ghiringhelli, N. Barišić, M. K. Chan, C. J. Dorow, G. Yu, X. Zhao, B. Keimer, and M. Greven. "Charge order and its connection with Fermi-liquid charge transport in a pristine high-$T_c$ cuprate," *Nat. Comm.* **5**, 5875 (2014).

[24] S.-C. Zhang, "Recent developments in the SO(5) theory of high $T_c$ superconductivity," *J. Phys. Chem. Solids* **59**, 1774 (1998).

[25] A. Himeda, T. Kato, and M. Ogata, "Stripe states with spatially oscillating d-wave superconductivity in the two-dimensional t-t'-J model," *Phys. Rev. Lett.* **88**, 117001 (2002).

[26] E. Berg, E. Fradkin, E.-A. Kim, S. A. Kivelson, V. Oganesyan, J. M. Tranquada, and S. C. Zhang, "Dynamical layer decoupling in a stripe-ordered high-$T_c$ superconductor," *Phys. Rev. Lett.* **99**, 127003 (2007).

[27] L. E. Hayward, D. G. Hawthorn, R. G. Melko, and S. Sachdev, "Angular fluctuations of a multicomponent order describe the pseudogap of $YBa_2Cu_3O_{6+x}$," *Science* **343**, 1336 (2014).

[28] W. E. Lawrence and S. Doniach, in Proc. 12th International Conference on Low Temperature Physics, Kyoto (1970), edited by E. Kanda (Keigaku, Tokyo, 1971).

[29] S. Doniach and M. Inui, "Long-range Coulomb interactions and the onset of superconductivity in the high-$T_c$ materials," *Phys. Rev. B* **41**, 6668 (1990).

[30] M. Tachiki, T. Koyama, and S. Takahashi, "Electromagnetic phenomena related to a low-frequency plasma in cuprate superconductors," *Phys. Rev. B* **50**, 7065 (1994).

[31] L. N. Bulaevskii, M. Zamora, D. Baeriswyl, H. Beck, and J. R. Clem, "Time-dependent equations for the phase differences and a collective mode in Josephson-coupled layered superconductors," *Phys. Rev. B* **50**, 12831 (1994).

[32] D. van der Marel and A. Tsvetkov, "Transverse optical plasmons in layered superconductors," *Czech. J. Phys.* **46**, 3165 (1996).

[33] D. van der Marel and A. Tsvetkov, "Transverse-optical Josephson plasmons: Equations of motion," *Phys. Rev. B.* **64**, 024530 (2001).

[34] A. Pimenov, A. Loidl, D. Dulić, D. van der Marel, I. M. Sutjahja, and A. A. Menovsky, "Magnetic field dependence of the transverse plasmon in SmLa$_{0.8}$Sr$_{0.2}$CuO$_{4-\delta}$," *Phys. Rev. Lett.* **87**, 177003 (2001).

[35] A. D. LaForge, W. J. Padilla, K. S. Burch, Z. Q. Li, S. V. Dordevic, K. Segawa, Y. Ando, and D. N. Basov. "Interlayer electrodynamics and unconventional vortex state in YBa$_2$Cu$_3$O$_y$," *Phys. Rev. B* **76**, 054524 (2007).

[36] A. E. Koshelev, "Electrodynamics of the Josephson vortex lattice in high temperature superconductors," *Phys. Rev. B* **76**, 054525 (2007).

[37] D. Fausti, R. I. Tobey, N. Dean, S. Kaiser, A. Dienst, M. C. Hoffmann, S. Pyon, T. Takayama, H. Takagi, and A. Cavalleri. "Light-induced superconductivity in a stripe-ordered cuprate," *Science* **331**, 189 (2011).

[38] C. R. Hunt, D. Nicoletti, S. Kaiser, T. Takayama, H. Takagi, and A. Cavalleri, "Two distinct kinetic regimes for the relaxation of light-induced superconductivity in La$_{1.625}$Eu$_{0.2}$Sr$_{0.125}$CuO$_4$," *Phys. Rev. B* **91**, 020505(R) (2015).

[39] M. Först, R. I. Tobey, H. Bromberger, S. B. Wilkins, V. Khanna, A. D. Caviglia, Y.-D. Chuang, W. S. Lee, W. F. Schlotter, J. J. Turner, M. P. Minitti, O. Krupin, Z. J. Xu, J. S. Wen, G. D. Gu, S. S. Dhesi, A. Cavalleri, and J. P. Hill. "Melting of charge stripes in vibrationally driven La$_{1.875}$Ba$_{0.125}$CuO$_4$: Assessing the respective roles of electronic and lattice order in frustrated superconductors," *Phys. Rev. Lett.* **112**, 157002 (2014).

[40] S. Kaiser, C. R. Hunt, D. Nicoletti, W. Hu, I. Gierz, H. Y. Liu, M. Le Tacon, T. Loew, D. Haug, B. Keimer, and A. Cavalleri, "Optically induced coherent transport far above $T_c$ in underdoped YBa$_2$Cu$_3$O$_{6+\delta}$," *Phys. Rev. B* **89**, 184516 (2014).

[41] W. Hu, S. Kaiser, D. Nicoletti, C. R. Hunt, I. Gierz, M. C. Hoffmann, M. Le Tacon, T. Loew, B. Keimer, and A. Cavalleri, "Optically enhanced coherent transport in YBa$_2$Cu$_3$O$_{6.5}$ by ultrafast redistribution of interlayer coupling," *Nat. Mater.* **13**, 705 (2014).

[42] M. Först, A. Frano, S. Kaiser, R. Mankowsky, C. R. Hunt, J. J. Turner, G. L. Dakovski, M. P. Minitti, J. Robinson, T. Loew, M. Le Tacon, B. Keimer, J. P. Hill, A. Cavalleri, and S. S. Dhesi. "Femtosecond x rays link melting of charge-density wave correlations and light-enhanced coherent transport in YBa$_2$Cu$_3$O$_{6.6}$," *Phys. Rev. B* **90**, 184514 (2014).

[43] R. Mankowsky, A. Subedi, M. Först, S. O. Mariager, M. Chollet, H. T. Lemke, J. S. Robinson, J. M. Glownia, M. P. Minitti, A. Frano, M. Fechner, N. A. Spaldin, T. Loew, B. Keimer, A. Georges, and A. Cavalleri. "Nonlinear lattice dynamics as a basis for enhanced superconductivity in YBa$_2$Cu$_3$O$_{6.5}$," *Nature* **516**, 71 (2014).

[44] R. Höppner, B. Zhu, T. Rexin, A. Cavalleri, and L. Mathey. "Redistribution of phase fluctuations in a periodically driven cuprate superconductor," *Phys. Rev. B* **91**, 104507 (2015).

[45] Quasiparticle contributions are accounted for in the effective medium fits by adding a small additional Drude term to the $1-f$ volume fraction. Within the THz frequency window, this is equivalent to adding a constant contribution to $\sigma_1(\omega)$.

[46] C. C. Homes, T. Timusk, D. A. Bonn, R. Liang, and W. N. Hardy, "Optical properties along the c-axis of YBa$_2$Cu$_3$O$_{6+x}$, for x=0.5 → 0.95: evolution of the pseudogap," Physica C 254, 265 (1995).

[47] D. Haug, V. Hinkov, Y. Sidis, P. Bourges, N. B. Christensen, A. Ivanov, T. Keller, C. T. Lin, and B. Keimer. "Neutron scattering study of the magnetic phase diagram of underdoped YBa$_2$Cu$_3$O$_{6+x}$," *New. J. Phys.* **12**, 105006 (2010).

[48] V. Hinkov, D. Haug, B. Fauqué, P. Bourges, Y. Sidis, A. Ivanov, C. Bernhard, C. T. Lin, and B. Keimer. "Electronic liquid crystal state in the high-temperature superconductor $YBa_2Cu_3O_{6.45}$," *Science* **319**, 597 (2008).

[49] Y. Wang, L. Li, and N. P. Ong, "Nernst effect in high-$T_c$ superconductors," *Phys. Rev. B* **73**, 024510 (2006).

[50] P. M. C. Rourke, I. Mouzopoulou, X. Xu, C. Panagopoulos, Y. Wang, B. Vignolle, C. Proust, E. V. Kurganova, U. Zeitler, Y. Tanabe, T. Adachi, Y. Koike and N. E. Hussey. "Phase-fluctuating superconductivity in overdoped $La_{2-x}Sr_xCuO_4$," *Nat. Phys.* **7**, 455 (2011).

[51] K. M. Kojima, S. Uchida, Y. Fudamoto, and S. Tajima, "New Josephson plasma modes in underdoped $YBa_2Cu_3O_{6.6}$ induced by a parallel magnetic field," *Phys. Rev. Lett.* **89**, 247001 (2002).

[52] H. Němec, L. Fekete, F. Kadlec, P. Kŭzel, M. Martin, J. Mangeney, J. C. Delagnes, and P. Mounaix. "Ultrafast carrier dynamics in $Br^+$-bombarded InP studied by time-resolved terahertz spectroscopy," *Phys. Rev. B* **78**, 235206 (2008).

[53] J. C. Delagnes, P. Mounaix, H. Němec, L. Fekete, *et al*., "High photocarrier mobility in ultrafast ion-irradiated $In_{0.53}Ga_{0.47}As$ for terahertz applications," *J. Phys. D: Appl. Phys.* **42**, 195103 (2009).

[54] J. Zielbauer and M. Wegener, "Ultrafast optical pump THz-probe spectroscopy on silicon," *Appl. Phys. Lett.* **68**, 1223 (1996).

[55] M. C. Beard, G. M. Turner, and C. A. Schmuttenmaer, "Transient photoconductivity in GaAs as measured by time-resolved terahertz spectroscopy," *Phys. Rev. B* **62**, 15764 (2000).

# Supplementary: Dynamical decoherence of the light induced interlayer coupling in YBa$_2$Cu$_3$O$_{6+\delta}$

C. R. Hunt[1,4], D. Nicoletti[1], S. Kaiser[1,3], D. Proepper[3], T. Loew[3], J. Porras[3], B. Keimer[3], A. Cavalleri[1,2]

[1] Max Planck Institute for the Structure and Dynamics of Matter, Hamburg, Germany
[2] Department of Physics, Oxford University, Clarendon Laboratory, Oxford, United Kingdom
[3] Max Planck Institute for Solid State Research, Stuttgart, Germany
[4] Department of Physics, University of California Berkeley, Berkeley, California, USA

**S1. Calculation of the transient complex optical response**

The experiment was performed in reflection geometry, with the pump beam striking the sample at normal incidence and the THz probe *p*-polarized at 30°. The reflected probe field in the absence of excitation, $E(t)$, and the pump-induced changes to the field at each time delay $\tau$, $\Delta E(t,\tau)$, were measured via electro-optic sampling and then independently Fourier transformed to extract the changes in the complex reflection coefficient $\tilde{r}'$, $\Delta \tilde{E}(\omega,\tau)/\tilde{E}(\omega) \equiv (\tilde{r}' - \tilde{r}_0)/\tilde{r}_0$. The equilibrium $\tilde{r}_0$ was taken from a separate measurement (see section S2).

The THz field samples a crystal volume ~20 times greater than the 15 $\mu$m pump pulse (see Figure S1.1.A-B). In order to isolate the THz response of the excited region alone, the system was modeled as a single excited layer, with complex refractive index $\tilde{n}(\omega)$, on top of a region that remains in equilibrium, with refractive index $\tilde{n}_0(\omega)$.[1] The excited layer thickness was set to the pump penetration depth $d_p$. A cartoon of this configuration is shown in Figure S1.1.C as a light purple excited layer on top of the equilibrium bulk in grey.

For comparison, we also utilized a multilayer approach, which treats the surface as a stack of $m$ thin layers, $\Delta z$, with refractive index decaying to its bulk value, $\tilde{n}(\omega) = \tilde{n}_0(\omega) + \Delta \tilde{n}(\omega) e^{-m\Delta z/d_p}$ (see also Ref. 2 and 3). This is depicted in the cartoon as the dark purple region. Both methods produced virtually identical results, as shown in Figure S1.2 for YBCO 6.5 at 100 K.

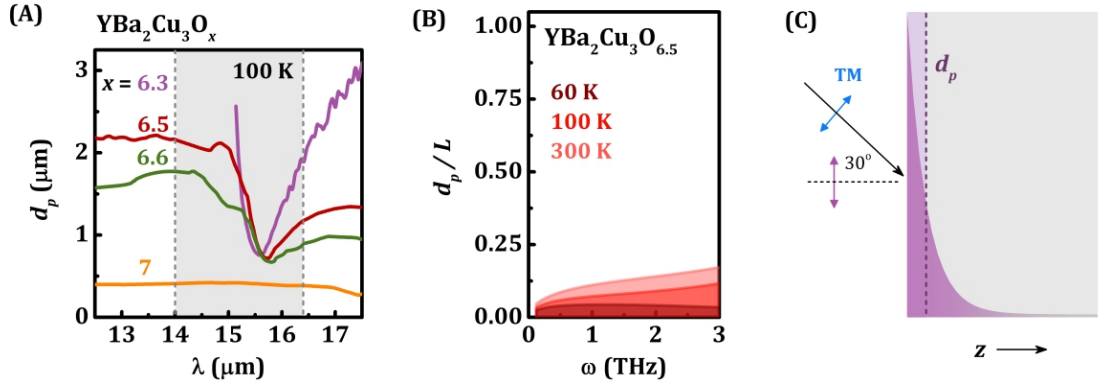

**Figure S1.1: Modeling the photo-excited crystal. (A)** Frequency-dependent penetration depth of YBCO 6.3-6.6 in the spectral region of the apical oxygen phonon. Dashed lines indicate the FWHM of the pump spectrum. **(B)** Pump-probe penetration depth mismatch for YBCO 6.5 at different temperatures. The pump penetration depth $d_p$ is taken at the minimum value at the phonon resonance. The frequency-dependent THz probe penetration depth is labeled $L$. **(C)** A cartoon depicting the excited sample and the pump-probe geometry. The pump impinges on the sample at normal incidence and excites a region at the sample surface. The photo-excited crystal is modeled as a single excited layer of thickness $d_p$ on an unperturbed bulk. This model agrees well with a second model that treats the excited layer as having a refractive index $\tilde{n}(\omega)$ at the surface that decays exponentially with depth $z$ to the equilibrium value $\tilde{n}_0(\omega)$.

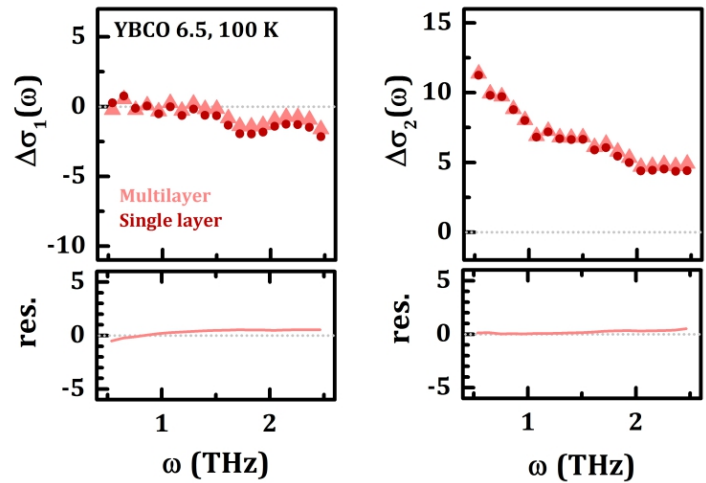

**Figure S1.2: Comparison of multilayer and single layer models.** The light-induced changes to the complex conductivity for YBCO 6.5 at 100 K, at $\tau = 0.8$ ps after excitation. The residual difference between the models is shown below the conductivity.

## S2. Equilibrium optical response of YBa$_2$Cu$_3$O$_{6.3}$

Since the *c*-axis optical response of YBCO at $T > T_c$ is flat and featureless in the 0.5-2.5 THz region, normalization of the transient spectra could be performed using literature data.[4,5,6] In addition, the *c*-axis equilibrium broadband dielectric function of YBCO 6.3 was measured on the same sample used for the pump-probe experiment, using far-infrared ellipsometry. These were then extended to $\omega \to 0$ using Drude-Lorentz fits (see Fig. S2.1). $\sigma_1(\omega \to 0)$ follows the high temperature dc resistivity trend, $\rho_0 \propto 1/T$, and the exponential doping dependence found in literature[7] (see Figure S2.2).

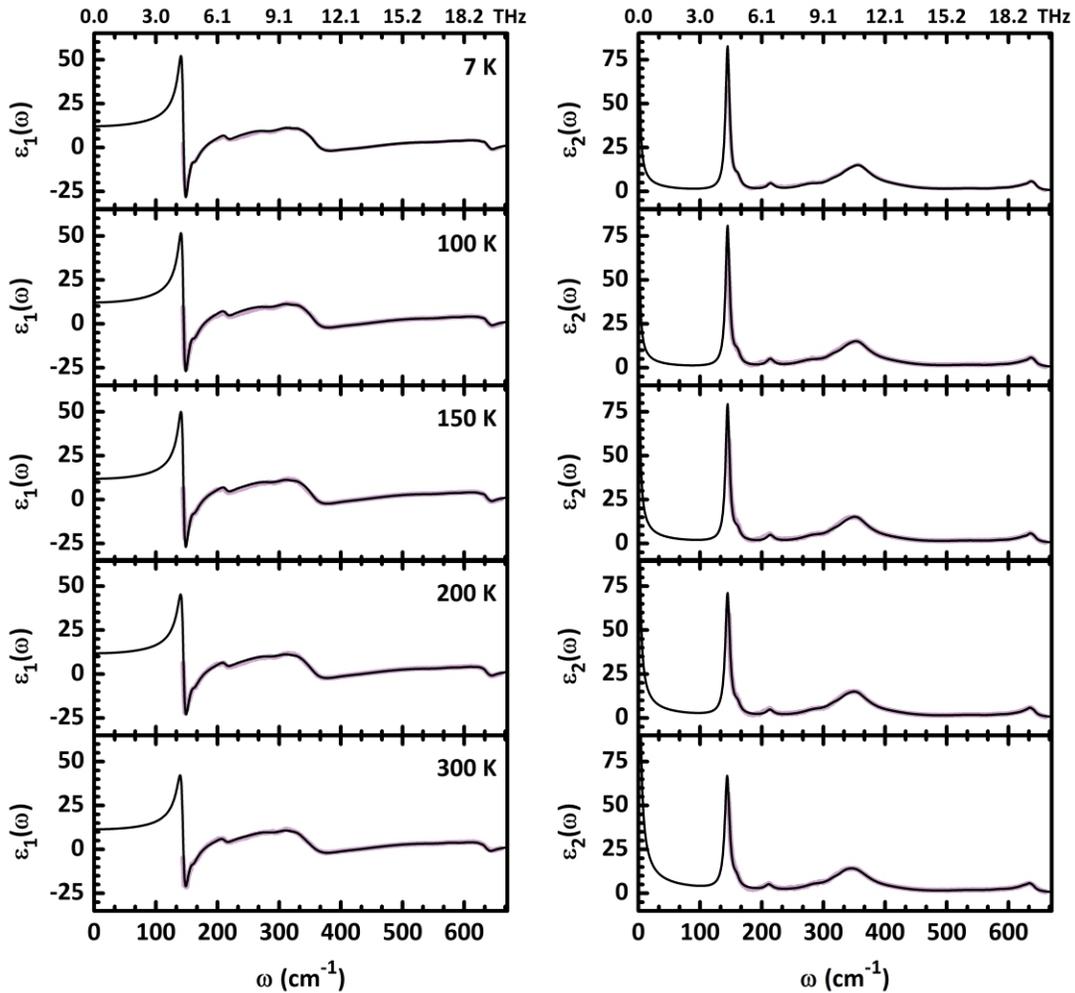

**Figure S2.1: Complex dielectric function of YBa$_2$Cu$_3$O$_{6.3}$.** Experimental data in purple, Drude-Lorentz fit in black.

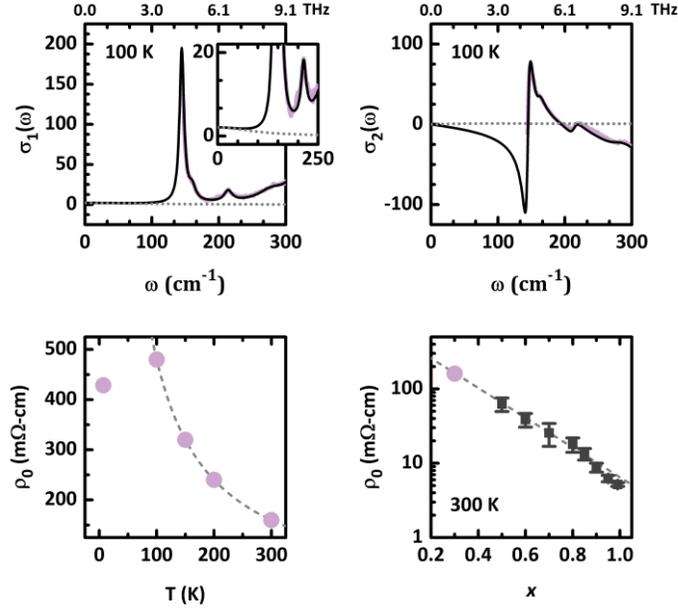

**Figure S2.2: Temperature and doping trends in the quasi dc conductivity. Top row:** The low frequency conductivity (in units $1/\Omega$-cm) calculated from the complex dielectric function, using $\varepsilon_\infty = 4.7$ (purple). Drude-Lorentz fits are shown in black. The Drude component is also plotted separately (dotted grey line). **Bottom row, left panel:** The estimated dc resistivity, $\rho_0 = 1/\sigma_1(\omega \to 0)$, following the trend $\rho_0 \propto 1/T$ for $T > 100$ K, as seen in literature for lower doped compounds. **Right panel:** Linear scaling of resistivity with doping. Black squares are data from Ref. 7.

### S3. The light-induced response in YBCO 6.3

Here in Fig. S3.1 we report the full complex conductivity of YBCO 6.3 at the peak of the response ($\tau = 0.8$ ps), at several temperatures (similar plots for YBCO 6.45-6.6 have already been reported in Ref. 2.) Black dashed lines indicate effective medium fits, as discussed in the main text.

Around 200 K, the best-fit photo-susceptible volume fraction $f$ approaches the percolation threshold, which is $f \geq 33\%$ for the Bruggeman form in Equation 2 in the main text (assuming spherical inclusions for the sake of simplicity). This response can also be fit with a single plasma mode, Eq. 1 in the main text, describing a material with metallic or superconducting inclusions with a volume fraction $f$ beyond the percolation threshold. Figure S3.2 shows a comparison between a single medium and an effective medium fit to the response at 200 K. Note that the small response in our THz frequency window, which cuts off above the plasma frequency, makes distinguishing these regimes difficult.

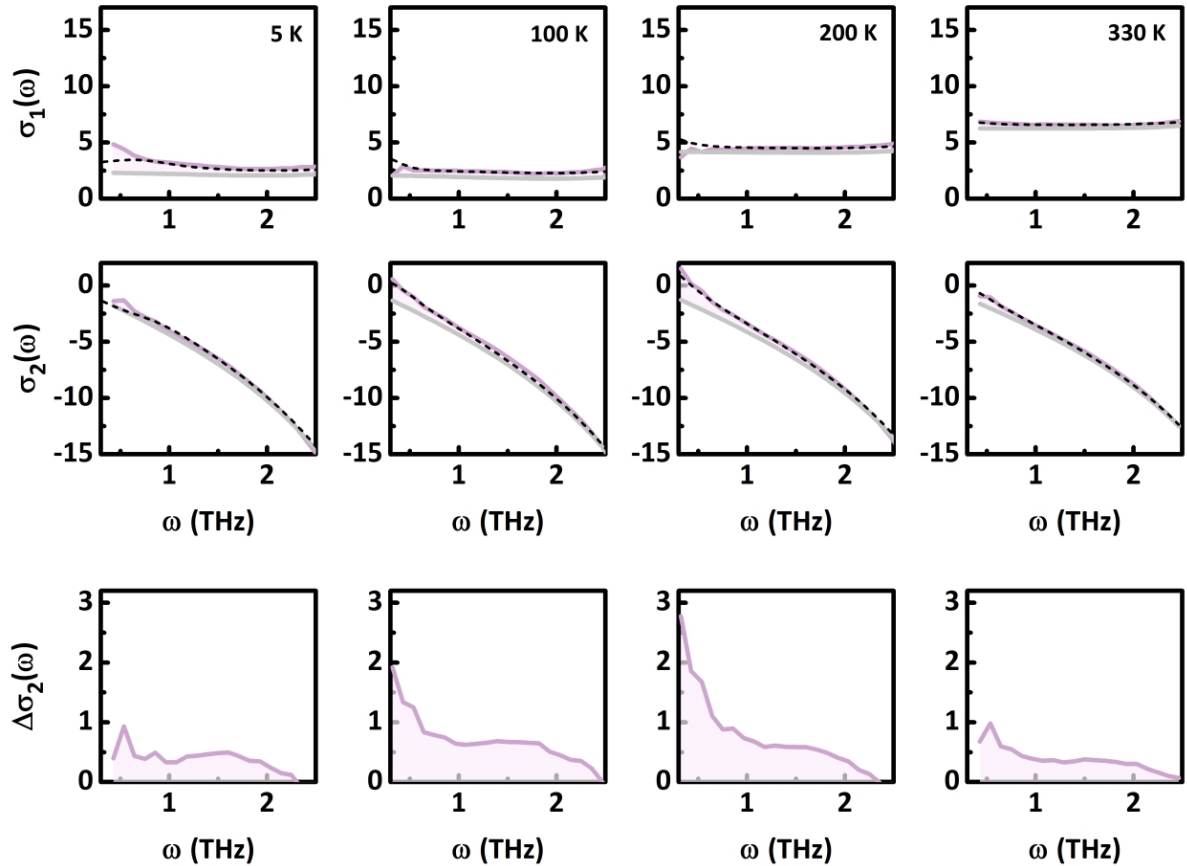

**Figure S3.1 Complex conductivity of photo-excited YBCO 6.3 at 5 K, 100 K, 200 K, and 330 K.** Real (top row) and imaginary (middle row) conductivity at $\tau = 0.8$ ps after excitation is shown in purple in units $1/\Omega$-cm. Effective medium fits appear as black dashed lines. Equilibrium conductivity is in grey. The light induced changes to the inductive conductivity, $\Delta\sigma_2(\omega) = \sigma_2(\omega, \tau = 0.8 \text{ ps}) - \sigma_{2,eq}(\omega)$, (bottom row) show a $1/\omega$-like increase to low frequency.

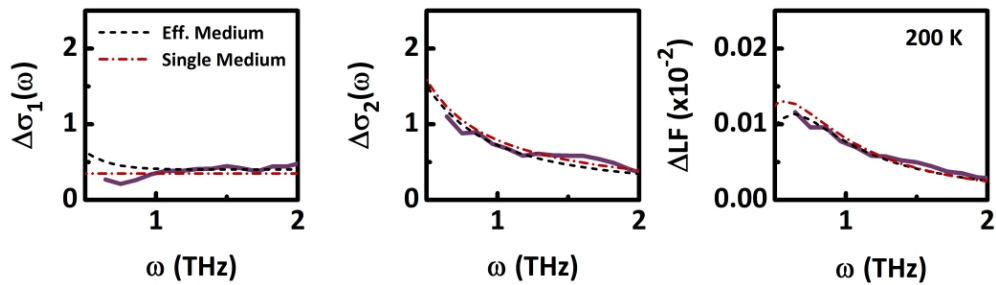

**Figure S3.2 Comparison of effective medium and single plasmon fits to YBCO 6.3 at 200 K.** Light-induced changes in the real (left) and imaginary (middle) conductivity and the loss function (right). Measured response is shown in purple, the single plasmon fits with black dashed lines, and the effective medium fit with grey dotted lines. The effective medium fit uses a photo-excited volume of $f = 32\%$.

## S4. Effect of pumping an in-plane phonon mode

An in-plane phonon mode of $B_{3u}$ symmetry also falls near 15 $\mu$m (see Fig. 3A). This is a Cu-O stretching mode in $CuO_2$ plane; analogous to the phonon resonantly excited in the experiments on lanthanides.[8] The planar lattice motions due to this mode are illustrated in Fig. S4B. Fig. S4C-D plots the changes in the real and imaginary conductivity of YBCO 6.5 at 300 K after excitation with 15-$\mu$m pulses polarized along the *a*-axis ($B_{3u}$ mode pumping) and *c*-axis ($B_{1u}$ mode pumping). No enhanced inductive response is observed after *a*-axis pumping; instead $\sigma_2(\omega)$ reduces. This result further highlights the importance of the apical oxygen mode excitation.

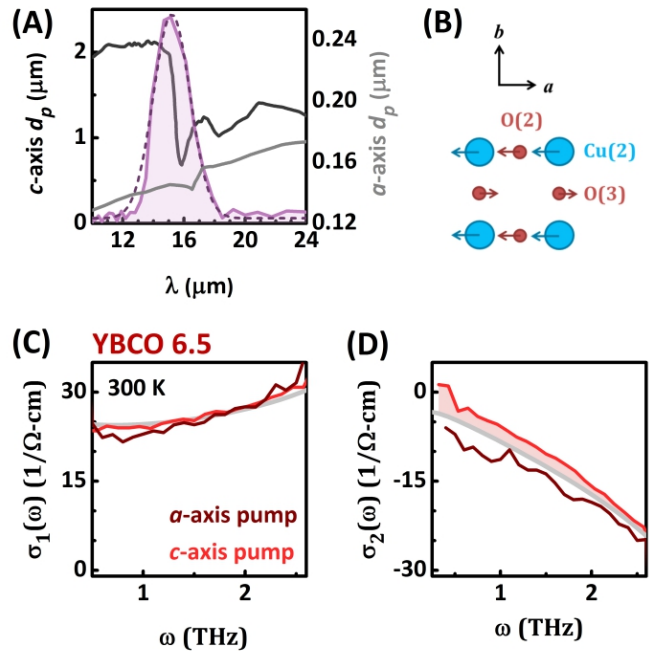

**Figure S4. Excitation of a $B_{3u}$ in-plane Cu-O stretching mode in YBCO 6.5. (A)** The pump penetration depth $d_p$ along the *c*-axis (black) and *a*-axis (grey). The kink in each profile indicates the frequency of the $B_{1u}$ and $B_{3u}$ phonon modes respectively. The pump spectrum is overlayed atop (purple) with a Gaussian fit (dashed line). The spectrum width is 1.5 $\mu$m. **(B)** A cartoon of the Cu (blue) and O (red) atoms in the $CuO_2$ planes. Arrows indicate the motion of the $B_{3u}$ mode. The bottom row plots the pump-induced changes to the **(C)** real and **(D)** imaginary conductivity probed along the *c*-axis. Both *c*-axis (red) and *a*-axis (dark red) pumping induce little change in $\sigma_1(\omega)$. The $\sigma_2(\omega)$ response is quite different, however, with only $B_{1u}$ (*c*-axis) mode excitation producing a positive change.

## S5. The time evolution of the inductive response

The relaxation of the light-induced plasma mode appears governed by a loss of coherence, which can be seen in terms of a suppression of the low frequency $\omega\Delta\sigma_2(\omega)$, which deviates from a constant behavior. Here we report an extensive analysis of the relaxation performed on YBCO 6.45, where the inductive response is large and the quasiparticle component is minimal at all temperatures.

Figure S5.1 shows $\omega\Delta\sigma_2(\omega)$ at different time delays, for several temperatures between 45 K and 300 K ($T_c = 35$ K). All data feature the same behavior. A flat $\omega\Delta\sigma_2(\omega)$ at early delays gives way first to a drop in the low frequency response below a frequency $\omega^*$. The downturn point $\omega^*$ shifts to higher frequency with time and is joined with an overall depression of the $\omega\Delta\sigma_2(\omega)$ level above $\omega^*$. The frequency-averaged $\langle\omega\Delta\sigma_2(\omega)\rangle$ above $\omega^*$ was extracted as a function of time at all temperatures and is plotted in Figure S5.2. The error bars reflect the standard deviation of $\langle\omega\Delta\sigma_2(\omega)\rangle$. The decay was fit with a double exponential to extract the timescales of the relaxation plotted in Fig. 4 of the main text.

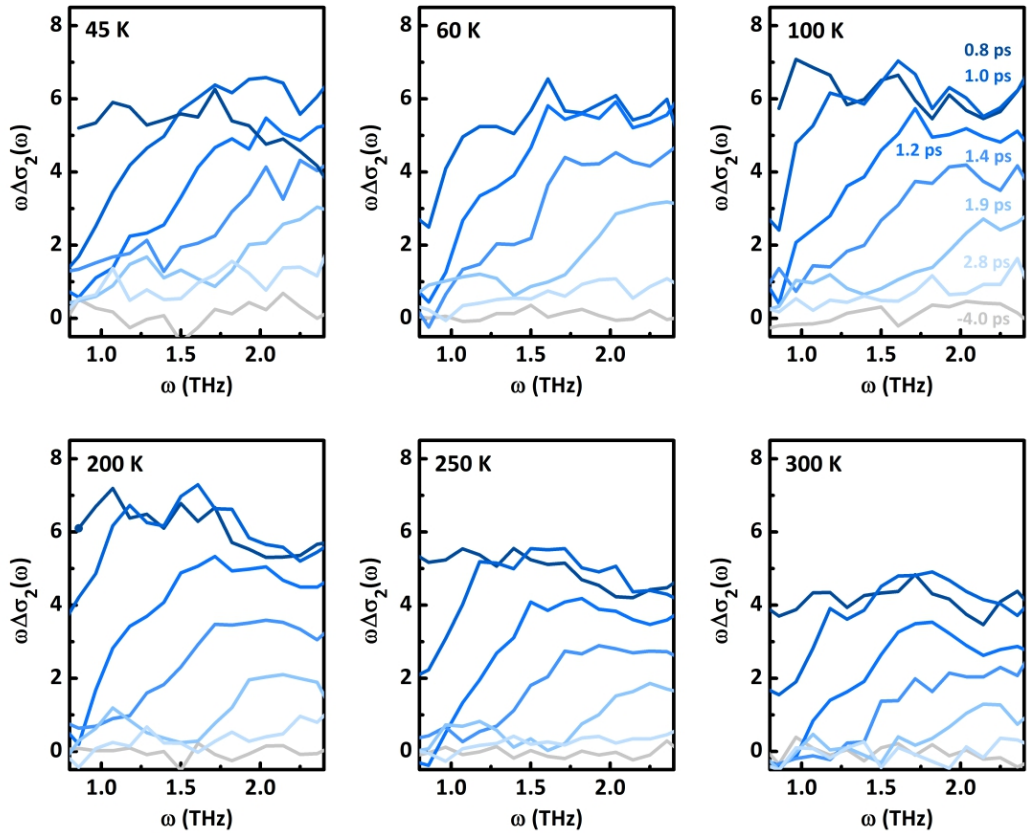

**Figure S5.1: The inductive response $\omega\Delta\sigma_2(\omega)$ for YBCO 6.45 at several temperatures between 45 K and 300 K.** Time delays are shown between 0.8 ps and 2.8 ps, with color code in the top-right panel.

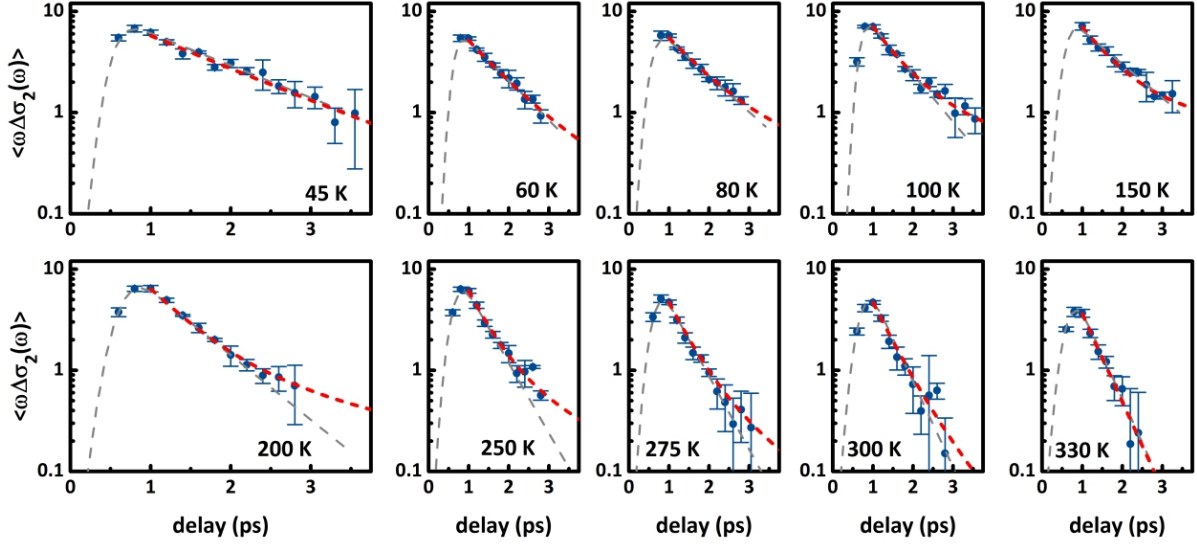

**Figure S5.2: The relaxation timescales of $\langle\omega\Delta\sigma_2(\omega)\rangle$.** The mean $\langle\omega\Delta\sigma_2(\omega)\rangle$, averaged where the response remains flat, is shown as a function of time for temperatures between 45 K and 330 K. The grey dashed lines indicate fits to a single exponential, plus error function to capture the behavior around the peak of the response. The red dashed lines are double exponential fits which more accurately capture the relaxation on longer time scales. The weight of each exponential component was held constant, with $\omega\Delta\sigma_2(\omega) \propto 0.8e^{-t/\tau_1} + 0.2e^{-t/\tau_2}$.

## S6. The splitting of the light-induced plasma mode

Fits to the optical conductivity attribute the drop in $\langle\omega\Delta\sigma_2(\omega)\rangle$ to a decrease in the volume fraction $f$ rather than a drop in the plasma frequency $\omega_p$. Figure S6.1A shows the loss function extracted from fits to the transient response, $\text{Im}(-1/\tilde{\varepsilon}_a)$ where $\tilde{\varepsilon}_a = \tilde{\varepsilon}_p$ at early time delays (see Eq. 1, main text) and $\tilde{\varepsilon}_a = \tilde{\varepsilon}_{2p}$ at later time delays (see Eq. 3, main text). The plasma frequency $\omega'_p = \omega_p/\sqrt{\varepsilon_\infty}$ is plotted as a function of time in panel S6.1B, showing the splitting of the plasmon at later delays. The filling fraction $f$ (Fig. S6.1C) relaxes with time following the double exponential behavior similar to that seen in $\langle\omega\Delta\sigma_2(\omega)\rangle$.

The splitting of the light-induced plasma mode is seen at all dopings and across all temperatures measured. Figure S6.2 shows the plasma mode for all four underdoped compounds, presented as a peak in the loss function $\Delta LF = \text{Im}(-1/\tilde{\varepsilon}) - \text{Im}(-1/\tilde{\varepsilon}_{eq})$, at two time delays after excitation. At 1 ps, the single peak in the response corresponds to a single plasma mode. By 1.8 ps after excitation, the peak has split into two peaks separated by ~1 THz, indicating two plasma modes.

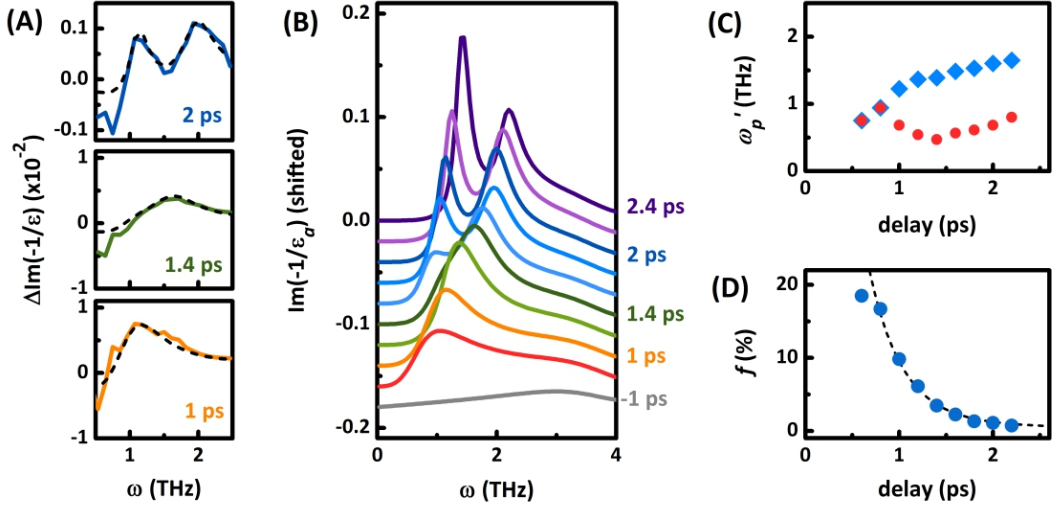

**Figure S6.1: Time evolution of the light-induced plasma mode in YBCO 6.45 at 200 K.** **(A)** The light-induced changes to the loss function $\Delta\mathrm{Im}(-1/\tilde{\varepsilon}) = \mathrm{Im}(-1/\tilde{\varepsilon}) - \mathrm{Im}(-1/\tilde{\varepsilon}_{eq})$ at three time delays after excitation. Dashed lines are effective medium fits. **(B)** The loss function $\mathrm{Im}(-1/\tilde{\varepsilon}_a)$ of a single plasma mode ($\tau \lesssim 1.0$ ps) or split plasma mode ($\tau \gtrsim 1.0$ ps) used in the fits ($\tilde{\varepsilon}_a = \tilde{\varepsilon}_p$ in Eq. 1 and $\tilde{\varepsilon}_a = \tilde{\varepsilon}_{2p}$ in Eq. 3 in the main text, respectively). **(C)** The reduced plasma frequency $\omega_p' = \omega_p/\sqrt{\varepsilon_\infty}$ as a function delay after excitation. **(D)** The volume fraction $f$ of the material with response $\tilde{\varepsilon}_a$.

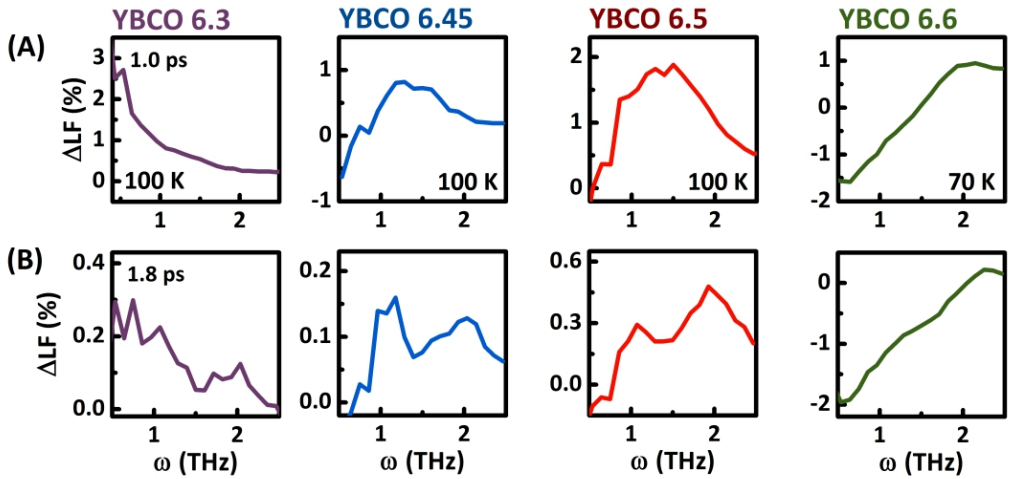

**Figure S6.2: Splitting of the light-induced plasma mode.** The transient loss function is shown at two time delays, 1.0 ps (top row) and 1.8 ps (bottom row). At early times, a single peak sits at the plasma mode frequency. At later times, the peak splits, indicating two modes.

[1] M. Born and E. Wolf, *Principles of Optics*. (Cambridge University Press, Cambridge, 1999) 7th Edition, p. 54-64.

[2] S. Kaiser, C. R. Hunt, D. Nicoletti, W. Hu, I. Gierz, H. Y. Liu, M. Le Tacon, T. Loew, D. Haug, B. Keimer, and A. Cavalleri, "Optically induced coherent transport far above $T_c$ in underdoped $YBa_2Cu_3O_{6+\delta}$," *Phys. Rev. B* **89**, 184516 (2014).

[3] C. R. Hunt, Ph.D. thesis, University of Illinois at Urbana-Champaign, 2015.

[4] C. C. Homes, T. Timusk, D. A. Bonn, R. Liang, and W. N. Hardy. "Optical properties along the c-axis of $YBa_2Cu_3O_{6+x}$ for $x$ = 0.50 → 0.95 Evolution of the pseudogap" *Physica C* **254**, 265-280 (1995).

[5] T. Timusk, C. C. Homes. "The role of magnetism in forming the *c*-axis spectral peak at 400 cm$^{-1}$ in high temperature superconductors," *Sol. State Comm.* **126**, 63-69 (2003).

[6] C. C. Homes, T. Timusk, D. A. Bonn, R. Liang, and W. N. Hardy. "Optical phonons polarized along the c-axis of $YBa_2Cu_3O_{6+x}$ for $x$ = 0.50 → 0.95," *Can. J. Phys.* **73**, 663-675 (1995).

[7] C. C. Homes, S. V. Dordevic, D. A. Bonn, R. Liang, W. N. Hardy, and T. Timusk. "Coherence, incoherence, and scaling along the c axis of $YBa_2Cu_3O_{6+x}$," *Phys. Rev. B* **71**, 184515 (2005).

[8] D. Fausti, R. I. Tobey, N. Dean, S. Kaiser, *et al.*, "Light-induced superconductivity in a stripe-ordered cuprate," *Science* **331**, 189 (2011); C. R. Hunt, D. Nicoletti, S. Kaiser, T. Takayama, H. Takagi, and A. Cavalleri, "Two distinct kinetic regimes for the relaxation of light-induced superconductivity in $La_{1.625}Eu_{0.2}Sr_{0.125}CuO_4$," *Phys. Rev. B* **91**, 020505(R) (2015); M. Först, R. I. Tobey, H. Bromberger, S. B. Wilkins, *et al.*, "Melting of charge stripes in vibrationally driven $La_{1.875}Ba_{0.125}CuO_4$: Assessing the respective roles of electronic and lattice order in frustrated superconductors," *Phys. Rev. Lett.* **112**, 157002 (2014).